\documentclass[12pt]{article}
\usepackage[polish,english]{babel}
\usepackage[latin1]{inputenc}
\usepackage{times}
\usepackage[T1]{fontenc}
\usepackage{epsfig}

\newcommand{\p}{\partial}

\begin{document}

\title{Signum-Gordon wave equation and its self-similar solutions }

\author{H. Arod\'z, $\;\;$ P. Klimas $\;$ and $\;$ T. Tyranowski \\$\;\;$ \\  Institute of Physics,
Jagiellonian University, \\ Reymonta 4, 30-059 Cracow, Poland}

\date{$\;$}

\maketitle

\begin{abstract}
We investigate self-similar solutions of evolution equation of a (1+1)-dimensional field model with the V-shaped
potential $U(\varphi) = | \varphi |,$  where $\varphi$ is a real scalar field. The equation contains a nonlinear
term of the form $sign(\varphi)$, and it possesses a scaling symmetry. It turns out that there are several
families of the self-similar solutions with qualitatively different behaviour. We also discuss a rather
interesting example of evolution with non self-similar initial data - the corresponding solution contains a
self-similar component.
\end{abstract}

\vspace*{2cm} \noindent PACS: 05.45.-a, 03.50.Kk, 11.10.Lm \\
\noindent Preprint TPJU - 15/2006

\pagebreak

\section{ Introduction}

Signum-Gordon equation\footnote{We thank Benny Lautrup from NORDITA for suggesting this name. } has the form
\begin{equation}
\frac{\partial^2 \varphi(x, t)}{\partial t^2} - \frac{\partial^2 \varphi(x, t)}{\partial x^2} = -
sign(\varphi(x, t)),
\end{equation}
where $\varphi$ is a real scalar field, and $x, t$ are dimensionless variables obtained by appropriate
rescalings of the physical position and time coordinates.   It follows from the Lagrangian
\[ L = \frac{1}{2} \partial_{\mu} \varphi \partial^{\mu}
\varphi - |\varphi|, \] where  $\mu =0, 1, \; \partial_0 = \partial_t, \; \partial_1 = \partial_x.$
 The $sign$ function has
the values $\pm 1$ when $ \varphi \neq 0,$ and 0 if $ \varphi = 0.$ Because of the $sign(\varphi)$ term, Eq. (1)
is nonlinear in a rather interesting way.  The corresponding field potential $U(\varphi) = |\varphi|$ has the
minimum at $\varphi =0$, and the field $\varphi$ can oscillate around the equilibrium value $\varphi =0$, but
Eq. (1) cannot be linearized even if  amplitude of the oscillations is arbitrarily small.

The signum-Gordon model has a rather sound physical justification. It has been obtained in a continuum limit (or
in a large wavelength approximation) to  certain  easy to built mechanical system with a large number of degrees
of freedom, see \cite{1, 2} for details. The model can also be applied in a description of  pinning of an
elastic string, which represents a vortex, to a rectilinear impurity in the case both the string and impurity
lie in one plane. Yet another interesting application is to  dynamics of a system of two global strings on a
plane. Such strings are represented by the real functions $\psi_1(x,t), \psi_2(x,t),$ and in certain
approximations their dynamics  is summarized in the following Lagrangian
\[
L_s(\psi_1, \psi_2) = \frac{1}{2} \partial_{\mu}\psi_1
\partial^{\mu}\psi_1 +  \frac{1}{2} \partial_{\mu}\psi_2\partial^{\mu}\psi_2
- V(\psi_1 - \psi_2),
\]
where $V$ is an extrapolation  of the well-known logarithmic interaction potential between separated global
strings to small values of $|\psi_1 - \psi_2|$:
\[
 V(\psi_1 - \psi_2) = a \ln\left( 1 + \frac{|\psi_1 - \psi_2|}{a}\right),
 \]
where $a$ is a positive constant (see  \cite{3} for a detailed discussion of the inter-string potential).  For
small $|\psi_1 - \psi_2|/a$ we have
\[ V(\psi_1 - \psi_2) \approx |\psi_1 - \psi_2|.
\] In this case the Euler-Lagrange equations for $ \psi_1, \psi_2$ obtained from Lagrangian $L_s$ imply that $ \varphi =\frac{1}{2}(
\psi_1 - \psi_2)$ obeys  the signum-Gordon equation. The solutions presented in our paper describe particular
cases of time evolution of all those systems. The present paper, however, is focused on  self-similar solutions
of the signum-Gordon equation rather than on its applications.

The signum-Gordon equation  possesses the exact scaling symmetry: if $\varphi(x, t)$ is its solution then
\begin{equation}
\varphi_{\lambda}(x, t) = \lambda^2 \: \varphi(\frac{x}{\lambda}, \frac{t}{\lambda})
\end{equation}
is a solution too, for any constant $\lambda >0$  \footnote{ Transformations of the form (2) with $\lambda <0$
are obtained as a product of the scaling transformation (2) with the reflections $ x \rightarrow -x, \; t
\rightarrow -t.$ Eq. (1) is invariant with respect to such reflections, and also with respect to 1+1 dimensional
Poincar\'e transformations.}.  This symmetry is one of the most interesting features of the model. As always
when there is a symmetry, one may search for solutions which are invariant under the symmetry transformations.
In the case of scaling they are the self-similar ones. In paper \cite{2} an example of such solutions of the
signum-Gordon equation has been given. Self-similar solutions of nonlinear equations play important role in
nonlinear dynamics. They have plenty applications - it is impossible to list them here. Beautiful presentation
of the self-similarity with its applications is given in \cite{4}. A shorter introdunction to this topic can be
found in \cite{5}.

The present paper is devoted to a thorough investigation of self-similar solutions  of the signum-Gordon
equation. Our main finding is that, rather surprisingly, there are quite many classes of such solutions. The
solutions presented in \cite{2} form merely a measure zero subset in just one such class: the segment $ S_0 =0,
\; -1/2 < \dot{S}_0 < 1/2$ in Fig. 1.  The various types of solutions differ in particular by the number of
isolated zeros of the field $\varphi$: they can have infinitely many, just one or no isolated zeros. Funnily
enough, all self-similar solutions are composed of quadratic polynomials in $x$ and $t$.

We have also investigated the evolution from certain particular initial data which are not self-similar, with
$\partial_x\varphi$ discontinuous at certain point.  While the late time behaviour of the corresponding solution
remains mysterious, we have noticed a very interesting phenomenon which occurs during the early stages of the
evolution. At first, the point of the discontinuity of $\p_x\varphi$ moves with a constant velocity equal to
$-1$ (`the velocity of light') until $\p_x\varphi$ becomes a continuous function of $x$. From this moment on,
the evolution of $\varphi$ locally, in a vicinity of the former point of discontinuity, follows a self-similar
solution with one isolated zero. This behaviour is universal in the sense that it does not depend on  details of
the initial data. Solutions of this class might be relevant for the description of the process of coalescence of
two strings, see Fig. 9.

All our analytic solutions have been checked by comparisons with purely numerical solutions of Eq. (1).
Actually, there was a very fruitful interplay of the two methods: certain solutions were  seen first numerically
and later obtained in the analytic form, whereas with others it was the other way round.

The plan of our paper is as follows. In Section 2 we discuss the Ansatz and initial data for the self-similar
solutions, and we present a map of such solutions. Section 3 is devoted to a detailed presentation of all
classes of the self-similar solutions: in subsections 3.1-3.5 we give explicit forms of all solutions.  In
Section 4 we consider the initial stages of evolution in the particular case of non self-similar initial data.
Section 5 contains a summary and remarks.

\section{The Ansatz and initial data } The physical context in which the signum-Gordon
equation (1) appears implies that the most interesting solutions are such that $\varphi(x, t)$ and their first
order partial derivatives $\partial_t\varphi, \;
\partial_x\varphi$ are continuous functions on the $(x, t)$ plane, perhaps except across the lines
$x= x_0 \pm t$ (the characteristics).  For example, a discontinuity of $\partial_t\varphi$ as a function of time
$t$ at a certain instant $t_1$  would mean that the velocity of a certain part of a continuous mechanical system
(e.g., of a string) suddenly jumps in spite of the fact that the force, given by the $sign(\varphi)$ term, is
always finite. Obviously, such discontinuities are unphysical.   On the other hand,  Eq. (1) implies that at
least one of the second order derivatives $\partial^2_t\varphi, \; \partial_x^2\varphi$ has to be discontinuous
when the piecewise constant function $F(x,t) = sign(\varphi(x,t))$ changes its value from 0 to $\pm1$.  At such
a point $\varphi =0,$ and in a certain neighbourhood of it $\varphi>0$ or $\varphi <0$. Then,  the left and
right second derivatives of $\varphi$ can have different values - if this is the case then the ordinary second
derivative of $\varphi$ does not exist, strictly speaking. Nevertheless, such solutions are physically
admissible. The proper mathematical framework for discussing them is based on the notion of weak solutions, see,
e.g. \cite{6}.

In the case of transformation (2) the  scale invariant Ansatz can be taken in the form
\begin{equation}
 \varphi(x, t) = x^2 S(y), \;\;\; y = \frac{t}{x}.
\end{equation}
We will consider solutions of Eq. (1) for $t > 0$ with initial data specified at $ t=0$. Solutions for $ t <0$
can be easily obtained with the help of the time reflection symmetry. It will turn out that $S(y)$ can be a
quadratic, linear or constant function of  $y$. Therefore, $\varphi$ will be continuous at $x=0$ in spite of the
fact that the scale invariant variable $y$ is singular at that point. The equivalent Ansatz
\[
\varphi(x,t) = t^2 \:T(\frac{x}{t})
\]
avoids the superficial singularity at $x=0$, but it is less convenient for incorporating the initial data.

The self-similar initial data have the form
\begin{equation}
\varphi(x,0) = \left\{ \begin{array}{ccc} R_0 x^2 & \mbox{for} & x \leq 0, \\ S_0 x^2 & \mbox{for} & x \geq 0,
\end{array} \right. \;\;\;
\left. \p_{t} \varphi(x, t)\right|_{t=0}= \left\{ \begin{array}{ccc} \dot{R}_0 x & \mbox{for} & x \leq 0, \\
\dot{S}_0 x & \mbox{for} & x \geq 0,
\end{array} \right.
\end{equation}
where $R_0, \dot{R}_0, S_0, \dot{S}_0$ are constants.  Such initial data are self-similar  because the point
$x=0$ is not shifted by the rescaling $ x \rightarrow x/\lambda$. In general, such  $\varphi(x,0)$ is of the
$C^1(R)$ class, while $\left.\p_t\varphi(x,t)\right|_{t=0}$  is of the class $C^0(R)$.

We shall restrict our considerations to the  case $R_0  = \dot{R}_0 =0$.  The other choice: $S_0 = \dot{S}_0 =0$
and $R_0, \dot{R}_0$ arbitrary is related to the previous one by the spatial reflection $ x \rightarrow -x,$
which is a symmetry of Eq. (1). Thus, in the main part of our paper we assume that
\begin{equation}
\varphi(x,0) = \left\{ \begin{array}{ccc} 0 & \mbox{for} & x \leq 0, \\ S_0 x^2 & \mbox{for} & x \geq 0,
\end{array} \right. \;\;\;
\left. \p_{t} \varphi(x, t)\right|_{t=0}= \left\{ \begin{array}{ccc} 0 & \mbox{for} & x \leq 0, \\ \dot{S}_0 x &
\mbox{for} & x \geq 0.
\end{array} \right.
\end{equation}
Such initial data  correspond to the following conditions for the function $S(y)$:
\begin{equation}
\lim_{y \rightarrow 0+} S(y) = S_0, \; \lim_{y \rightarrow 0-} S(y) = 0, \\
\lim_{y \rightarrow 0+} S'(y) = \dot{S}_0, \; \lim_{y \rightarrow 0-} S'(y) = 0.
\end{equation}
Let us remark that solutions for more general self-similar initial data (4) can be obtained in certain cases
just by combining solutions obeying (5) with the ones obtained from them by applying the spatial reflection, but
there are also cases in which such approach does not work.

The Ansatz (3) reduces Eq. (1)  to the following ordinary differential equation for the function $S(y)$
\begin{equation}
 (1-y^2)\: S'' +2y\:S' -2\:S = - sign(S),
\end{equation}
where $ S' = d S / d y$, and the $sign(S)$ function has the values $+1$ or $-1$ for $S>0$ or $S<0$,
respectively, and $sign(0) =0.$ Notice that at the points $y = \pm 1$ (which correspond to characteristics of
Eq. (1)) the coefficient in front of the second derivative term in Eq. (7) vanishes. This has the consequence
that the first derivative $ dS/ dy $ of the solution does not have to be continuous at these points. Below we
shall make use of this possibility. In order to obtain weak solutions of this equation we first solve it
assuming that $S > 0$ or $S < 0$ or $S = 0$. Such partial solutions have a rather simple form of quadratic
polynomials in $y$:
\[\mbox{when}
\;\;\; S > 0: \;\; \;\;\;\; S_+(y) =  - \frac{\beta}{2}  (y^2 +1) + \frac{\alpha}{2} y + \frac{1}{2},
\]
\[
\mbox{when} \:\;\; S < 0:\;\;\;\;\;\; S_-(y) =   \frac{\beta'}{2} (y^2 +1) - \frac{\alpha'}{2} y - \frac{1}{2},
\]
where $\alpha, \beta, \alpha', \beta'$ are arbitrary constants. The polynomial solutions are valid on
appropriate intervals of the $y$-axis determined by the conditions $S
> 0$ or $S < 0$, respectively. There is also the trivial solution
\[ S_0(y) = 0. \]

 Next, we match such partial solutions requiring that $S$ and $S'$ are continuous functions
of $y,$  except at the point $y =  1,$ where only continuity of $S$ is required. It turns out that the
continuity of only $S$ at $y =  1$ is sufficient to ensure the implied by  wave equation (1) continuity of \[
\frac{\partial \varphi}{\partial x_+} = \frac{1}{2} \left(\frac{ \partial \varphi}{\partial x} + \frac{\partial
\varphi}{\partial t} \right), \]
 where $ x_+ = x + t,$  across  the characteristic line
$x=t.$ The matching conditions together with the initial data (5) determine the constants
 $\alpha, \beta, \alpha', \beta'$  from the partial solutions.
In this manner we have constructed exact, self-similar weak solutions of the Cauchy problem (5) for all values
of parameters $S_0, \dot{S}_0.$ Because $x \in (-\infty, \infty)$, the variable $y$ can take all real values.
The factor $x^2$ in (3) makes the solutions finite at $x = 0.$

Depending on the values of  $S_0, \dot{S}_0$ the self-similar solutions can have quite different forms. As one
can see from the `map' presented in   Fig. 1, the space of such solutions of the signum-Gordon equation is
surprisingly rich.

\begin{center}
\begin{figure}[tph!]
\hspace*{1cm} \includegraphics[height=10.5cm, width=12cm]{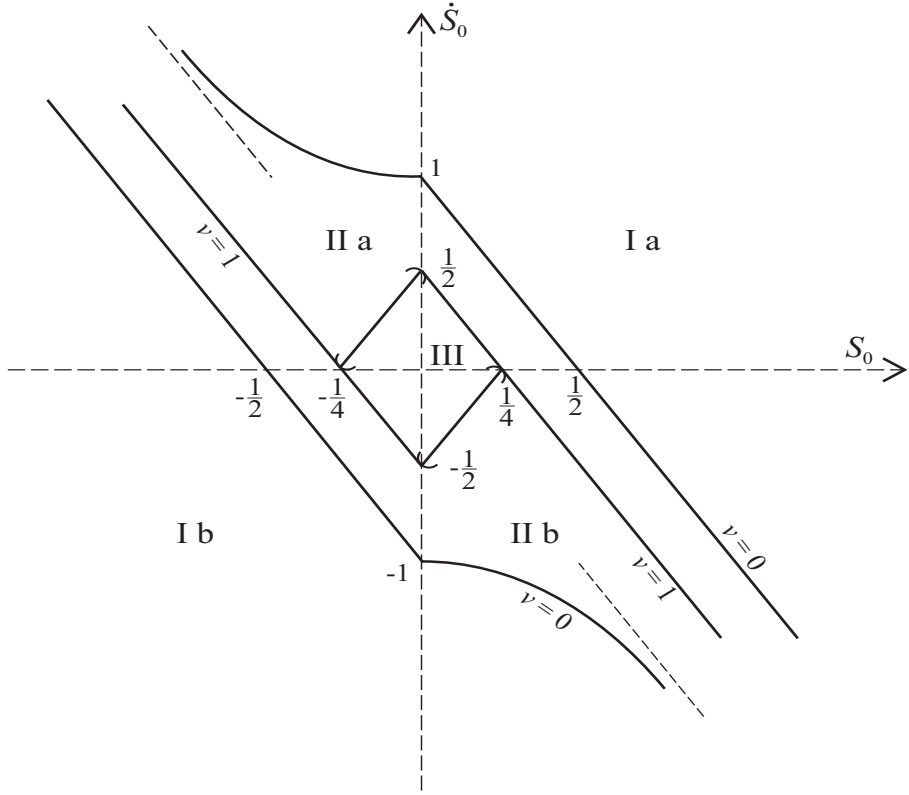} \caption{ The map of the self-similar
solutions of the signum-Gordon equation (1). Each point in the $(S_0, \dot{S}_0)$ plane represents one solution.
The symmetry of the picture with respect to the reflection $ (S_0, \dot{S}_0) \rightarrow  (- S_0, - \dot{S}_0)$
is related to the reflection symmetry $\varphi \rightarrow - \varphi$ of the signum-Gordon equation. Because of
this symmetry it is sufficient to discuss solutions from the half-plane $S_0 \geq 0$. The dashed lines are
auxilliary: two coordinate lines, and the two lines $\dot{S}_0 = - 2 S_0 $ which are asymptotically tangent to
the lines $v=0$. The continuous lines are border lines between classes of solutions, but their points also
represent some special solutions discussed in subsections 3.1 (the  $v=0, \: S_0 \geq 0$ rectilinear half-line),
3.2 (the $v=0, S_0 \geq 0$ curved line), 3.4 (the open segment connecting the point $(0, -1/2)$ with the point
$(1/4, 0)$), and 3.5 (the $v=1, S_0 \geq 0$ half-line). $v$ is a certain velocity defined in the text. The
lower, curved $v=0$ line is a cubic curve given by equation (16), the other one is obtained by the reflection.
The rounded brackets, which enclose two line segments - the edges of the rhomb III, indicate that these segments
do not contain ends - the end points belong to the two $v=1$ half-lines. }
\end{figure}
\end{center}

\section{ Self-similar solutions }
Below we give  analytic forms of the self-similar solutions with initial data (5). We assume that $S_0 \geq 0.$
Solutions with negative values of $S_0$ can be obtained by the reflection $ \phi(x,t) \rightarrow - \phi(x,t)$,
which in particular means that $S_0 \rightarrow - S_0, \; \dot{S}_0 \rightarrow - \dot{S}_0.$

\subsection{ Solutions of types: Ia, IIa, $v=0$ }
These solutions lie in the open region above the rhomb III and above the $v=1, \; S_0 \geq 0$ line, see Fig. 1.
The solutions are obtained by matching  the trivial solution with the partial solution $S_+$ at the point $y =
1/v$ which corresponds to the line $x =v t$. The velocity $v$ has values in the interval $(-1, 1)$, and it is
given by formula (8). It turns out that one has to use a second solution of the type $S_+$ which matches the
previous one at the point $y=1$  corresponding to the characteristic line $x=t$. That latter $S_+$ solution has
the form $ S_0 +  \dot{S}_0 y - (1/2 - S_0) y^2,$ where $S_0 \geq 0, \: \dot{S}_0$ are the parameters specifying
the initial data according to formula (5).

The velocity $v$ is given by the formula
\begin{equation}
v = \frac{1}{2 S_0 + \dot{S}_0} -1,
\end{equation}
which follows from the matching condition at the point $y=1$. In the region Ia the velocity has values from the
interval $(-1,0)$, while in the region IIa  $ v \in (0, 1). $ On the line $2 S_0 + \dot{S}_0 =1,$ which
separates the two regions, $v = 0$. The full solution has the form
\begin{equation}
\varphi(x,t) = \left\{ \begin{array}{lcl} \varphi_0 = 0  & \;\;\; \mbox{for} \;\;\; & x \leq vt, \\
\varphi_1 = \frac{(x-vt)^2}{2 (1-v^2)}&     \;\;\; \mbox{for} \;\;\;   &   vt \leq x \leq t, \\
\varphi_2 = S_0 x^2 + \dot{S}_0 t x + ( S_0 - \frac{1}{2}) t^2 &    \;\;\; \mbox{for} \;\;\;   &x \geq t.
\end{array} \right.
\end{equation}
It is depicted in Fig. 2.
\begin{center}
\begin{figure}[tph!]
\hspace*{1cm}
\includegraphics[height=6.0cm, width=12cm]{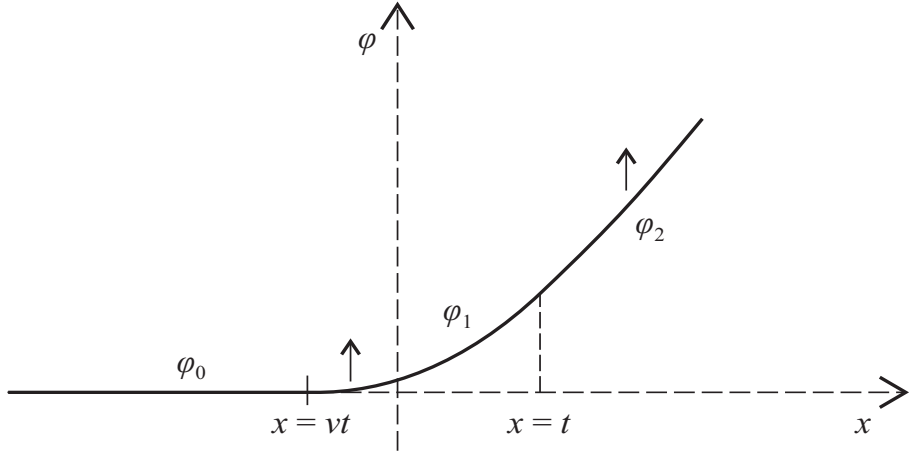}
\caption{Picture of a solution of type Ia  for which $-1 < v <0$. $v$ is the velocity of the point on the $x$
axis at which $\varphi_1$ merges with the trivial solution $\varphi =0$.   The two vertical arrows indicate that
values of $\varphi$ increase with time - the $\varphi >0$ part moves to the left. Solutions of the type IIa have
a similar shape, but they move to the right ($ 0 < v <1).$  }
\end{figure}
\end{center}

The half-line $ 2 S_0 + \dot{S}_0 =1, \; S_0 \geq 0, v=0, $  which separates the classes Ia and IIa, includes
the static solution
\begin{equation} \varphi_{s+}(x) = \left\{
\begin{array}{lcl}  0 & \;\;\; \mbox{for} \;\;\; &  x \leq 0, \\ x^2/2 & \;\;\; \mbox{for} \;\;\; & x\geq 0,
\end{array} \right.
\end{equation}
for which $ S_0 =1/2, \; \dot{S}_0 =0.$  Other solutions from that line are time dependent. They coincide with
the static one in the growing with time interval $ 0 \leq x \leq t,$ while at the points $x
> t$ they have the form $\varphi_2(x,t)$, see the second and third lines in formula (9). Asymptotically,
as $t$ grows to infinity, such  $\varphi$ entirely covers the static solution.

The signum-Gordon equation is invariant under Lorentz transformations
\[
t' = \gamma ( t - w x), \;\;\; x' = \gamma ( x - w t),
\]
where $ \gamma = 1 / \sqrt{ 1 - w^2}$, $ |w| <1.$ Such transformations preserve self-similarity of solutions.
One can see that the static solution (10) is transformed into the solution with $S_0 = \gamma^2/2, \; \dot{S}_0
= - w \gamma^2.$ In this case formula (8)  gives  $v=w.$ For these particular solutions the functions $
\varphi_1$ and $\varphi_2$ in formula (9) have the same form.

\subsection{ Solutions of types: Ib, IIb,  $v=0$. }
These solutions lie in the open region below the rhomb III and the $v=1, \: S_0 >0$ line, see Fig. 1. Solutions
of this type have the form shown in Fig. 3. Solutions from the half-line $S_0 =0,\; \dot{S}_0 \leq -1/2$ are not
considered here because they can be obtained from solutions lying on the $S_0 =0, \;\dot{S}_0 \geq 1/2$
half-line with the help of the reflection $ \phi \rightarrow - \phi.$
\begin{center}
\begin{figure}[tph!]
\hspace*{1cm}
\includegraphics[height=6.0cm, width=12cm]{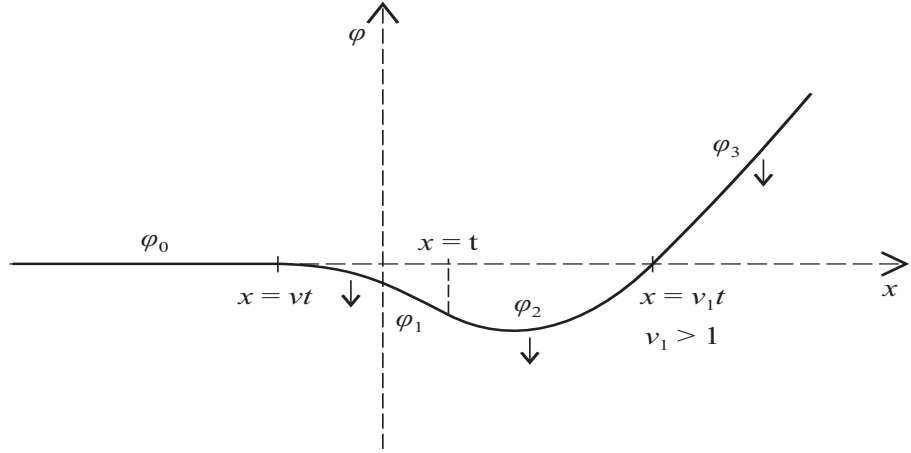}
\caption{Picture of a solution of type Ib  for which $-1 < v <0$. The  arrows indicate that values of $\varphi$
decrease with time - the $\varphi <0$ part expands in both directions.  Solutions of the type IIb have a similar
shape, but they move to the right ($ 0 < v <1).$ }
\end{figure}
\end{center}

They are composed of the trivial solution $S=0$ in the region $ y > 1/v
>1,$ one solution $S_+$ in the region $y < 1/v_1 < 1,$  and two solutions $S_-$ in the region $ 1/v_1 < y <
1/v$ which continuously match  each other at the point $y=1$. The pertinent formula for $\varphi$ has the
following form
\begin{equation}
\varphi(x,t) = \left\{ \begin{array}{lcl}
\varphi_0 = 0  & \;\;\; \mbox{for} \;\;\; & x \leq vt, \\
\varphi_1 = - \frac{(x-vt)^2}{2 (1-v^2)}&     \;\;\; \mbox{for} \;\;\;   &   v t \leq x \leq t, \\
\varphi_2 =  \frac{\beta_2}{2} (x^2 + t^2) - \frac{\alpha_2}{2} x t  - \frac{1}{2} x^2  &
   \;\;\; \mbox{for} \;\;\;   &    t \leq x \leq v_1 t, \\
\varphi_3 = S_0 x^2 + \dot{S}_0 t x + ( S_0 - \frac{1}{2}) t^2 &    \;\;\; \mbox{for} \;\;\;   &x \geq v_1 t.
\end{array} \right.
\end{equation}
Here $S_0, \dot{S}_0$ are given in the initial data (5). The velocity $v_1$ is obtained from the condition
$\varphi_3(v_1 t, t) =0$, and  $ \alpha_2, \beta_2$ from the matching conditions at $ x= v_1 t,$ i.e. at $y =
1/v_1 <1.$ Straightforward calculations give
\begin{equation}
v_1 = \frac{ - \dot{S}_0 + \sqrt{ \dot{S}_0^2 - 4 S_0^2 + 2 S_0}}{2 S_0},
\end{equation}
\begin{equation}
\alpha_2 =  2 \frac{v_1 + ( 1 + v_1^2)\sqrt{ \dot{S}_0^2 - 4 S_0^2 + 2 S_0}}{v_1^2 -1},
\end{equation}
\begin{equation}
\beta_2 =  \frac{v_1^2 + 2 v_1\sqrt{ \dot{S}_0^2 - 4 S_0^2 + 2 S_0}}{v_1^2 -1}.
\end{equation}
Finally, the velocity $v$ is obtained from the condition of continuity of $\varphi(x,t)$ at the point $x=t:$
\begin{equation}
v =  \frac{2 \beta_2 - \alpha_2}{2 + \alpha_2 - 2 \beta_2}.
\end{equation}
Formulas (12-15) imply that $ v_1 >1$ and $-1 <  v < 1$.

The condition $v=0$ is satisfied by $ (S_0, \: \dot{S}_0)$ such that
\begin{equation}
\eta = \frac{1}{2} + \frac{1}{2 \xi^2},
\end{equation}
where
\[ \eta = 2 S_0 - \dot{S}_0, \;\;  \xi = 2 S_0 + \dot{S}_0.
\]
These points form the lower curved line in Fig. 1. Solutions of this kind are related to another static solution
of Eq. (1), namely to
\begin{equation} \varphi_{s-}(x) = \left\{
\begin{array}{lcl}  0 & \;\;\; \mbox{for} \;\;\; &  x \leq 0, \\ - x^2/2 & \;\;\; \mbox{for} \;\;\; & x\geq 0,
\end{array} \right.
\end{equation}
for which $ S_0 = - 1/2, \; \dot{S}_0 =0.$      These  $v=0$ solutions are time dependent. Their $ \varphi_1$
part, see formula (11), coincides with the static one in the growing with time interval $ 0 \leq x \leq t,$
while at the points $x
> t$ their  form is given by $\varphi_2(x,t), \; \varphi_3(x,t).  $ Asymptotically,
as $t$ grows to the infinity, we recover the static solution.

\subsection{Solutions of type III }

Let us try to put together a number of the partial solutions $S_+, S_-$  in order to cover as large as possible
interval of the $y \geq 0$ half-axis, see Fig. 4,   where $S_k, \:k=1,2, \ldots,$ denote the consecutive partial
solutions $S_{\pm}$.

 The restriction to $y \geq 0,$ see Fig. 4,  is quite natural. The point is that $y = 0_-$ corresponds to $x= -
\infty,$ while $x=0_+$ to $x = +\infty$. Hence, there is no physical reason for demanding that $S(y)$ is
continuous at $ y=0$ -  the continuity at $y=0$ would mean that we arbitrarily impose the restriction $
\varphi(\infty, t) = \varphi(-\infty,t).$ Therefore, the solution $S_1$ is terminated at $y=0$, and in the
region $ y <0$ we take the trivial solution $ S_0(y) =0.$
 On the other hand, both $y = -\infty$ and $y = +\infty$ correspond to $x=0$, where the solution $\varphi$
is required to be a continuous and differentiable function of $x$.

\begin{center}
\begin{figure}[tph!]
\hspace*{1cm}
\includegraphics[height=4.5cm, width=12cm]{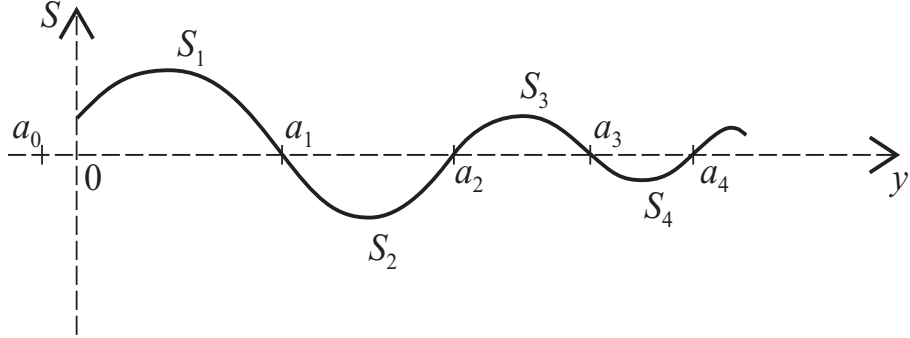}
\caption{The  polynomials $S_k$ and the matching points  $a_k, \; k=1, 2, \ldots .$ }
\end{figure}
\end{center}

Let us write the partial solutions in the form
\[
S_k(y) = \frac{1}{2} (-1)^k [ \beta_k(y^2+1) - \alpha_k y -1].
\]
Equation (1) implies that the following matching conditions at the points $y = a_k, k \geq 1,$ have to be
satisfied:
\begin{equation}
S_k(a_k) = 0 = S_{k+1}(a_k), \;\;\; S'_k(a_k) = S'_{k+1}(a_k),
\end{equation}
provided that $ a_k \neq \pm 1.$ When $y= \pm 1$, only continuity of $S(y)$ has to be required.

The matching conditions (18) yield the recurrence relations:
\begin{equation}
\alpha_{k+1} = \frac{4 a_k}{1 - a_k^2} - \alpha_k, \;\;\;\beta_{k+1} = \frac{2}{1 - a_k^2} - \beta_k,\;\;\;
a_{k+1} = \frac{2 a_k - (1 + a_k^2) a_{k-1}}{1 + a_k^2 - 2 a_k a_{k-1}},
\end{equation}
where $\alpha_1 = a_1 > 0, \;\; a_0 \leq 0, $ and
\begin{equation}
\beta_1= \frac{1}{1- a_0 a_1}, \;\; \alpha_1 = \frac{a_0 + a_1}{1 - a_0 a_1}.
\end{equation}
The relations (20) follow from the fact that $a_0, a_1$ are the zeros of the polynomial $S_1(y)$. It is clear
from Fig. 4 that $a_0 \leq 0, \; a_1 > 0.$ The polynomial $S_1(y)$ has zeros at $y=a_0, a_1.$

The solution is single-valued when $a_{k+1} \geq a_k.$ Simple calculations show that this condition taken for
$k=1, 2$ implies that $a_1 \leq 1,$  $ a_2 \leq 1, $ and that  $a_0 \geq -1$. Furthermore, we notice that $a_0 =
-1$ or $a_1=1,$  always give $ a_2 =1$, what implies that  $a_k =1$ for all $k>2$. Such solutions, i.e., with
$a_2=1,$ belong to the classes of solutions discussed in subsections 3.4, 3.5, except for the case $a_0 =-1, a_1
=0$ in which we obtain the trivial solution $S_0=0$.  Therefore, in the remaining part of this subsection we
assume that
\[ -1 < a_0 \leq 0,\; 0 < a_1 <1.
\]

Recurrence relations (19) have the following general solutions:
\begin{equation}
\alpha_k = \frac{1}{1+r} \left[\frac{1}{p r^{k-1}} - p r^k + (-1)^k (\frac{1}{p} - q ) \right] - (-1)^k
\alpha_1,
\end{equation}
\begin{equation}
\beta_k = \frac{1}{2} + \frac{1}{2(1+r)} \left[\frac{1}{p r^{k-1}} + p r^k  + (-1)^k (1 + \frac{1}{p})(1+q)
\right] - (-1)^k \beta_1,
\end{equation}
and
\begin{equation} a_k = \frac{p^{k-1}- q^k}{p^{k-1}+ q^k},
\end{equation}
where \begin{equation}  q = \frac{1 - a_1}{1+a_1}, \;\; p = \frac{1-a_0}{1+a_0}, \;\; r = \frac{q}{p}.
\end{equation}
Notice that $ 0 \leq q <1$ and $p >1.$

Solutions (21-23) have been conjectured after seeing several first iterations of recurrence relations (19), and
checked by substitution into these relations.  In paper \cite{6} we were able to find only solutions with $a_0
=0, \; a_1 <1. $ They lie on the segment $ S_0 =0, \; 0 < \dot{S}_0 < 1/2$ inside the rhomb III in Fig. 1.

Formula (23) implies that $ a_k <1,$ and that $  a_k \rightarrow 1 $ when $ k \rightarrow \infty.$ Therefore,
the piecewise polynomial solution constructed above covers only the interval $[0,\:1)$. For this reason we
introduce a special notation for it, namely $S_{pp}(y)$. Calculations show that $S_{pp}(y)$ vanishes  when $y
\rightarrow 1$. On the other hand, the first derivative $S_{pp}'(y)$ does not vanish in that limit, but it
remains finite.

This solution can be  extended  to the full range of $y$. We have already taken  the trivial solution $S_0(y) =
0$ in the region $ y < 0$. In  the region $y \geq 1$ we also take the trivial solution which is consistent with
continuity of $S(y)$ at $ y=1$. The first derivative $d S/ d y$ in general is not continuous at $y=1$, but this
is not forbidden by Eq. (7) because the coefficient $1-y^2$ in front of $S''$ vanishes when $y=1$. Notice that $
y = \pm 1 $ correspond to the characteristics $ x = \pm t$ of the signum-Gordon equation.

The solution we have just constructed has the following form:
\begin{equation}
\varphi(x,t) = \left\{ \begin{array}{lcl}
\varphi_0 =  0  & \;\;\; \mbox{for} \;\;\; & x \leq t, \\
\varphi_{pp} = x^2 S_{pp}(\frac{t}{x}) &     \;\;\; \mbox{for} \;\;\;   &   x >t.
\end{array} \right.
\end{equation}
Snapshot of the solution is presented in Fig. 5.

\begin{center}
\begin{figure}[tph!]
\hspace*{1cm}
\includegraphics[height=4.5cm, width=12cm]{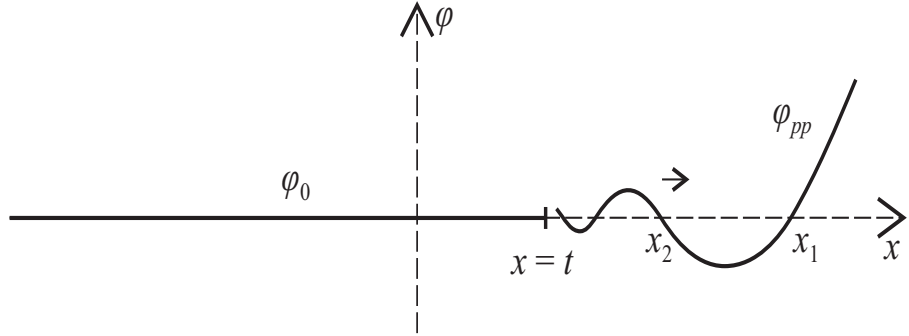}
\caption{ The self-similar solution of type III. The zeros $x_k = v_k t$ of $\varphi$ accumulate at $x=t$
because $v_k = 1/a_k,$ where $ a_k \rightarrow 1$ when $ k \rightarrow \infty.$  The arrow indicates that whole
structure moves to the right. }
\end{figure}
\end{center}

The zeros of $\phi$ are given by the formula  $x_k(t) = t / a_k.$ They move with the constant velocities $v_k =
1/a_k >1.$

The parameters $S_0, \dot{S}_0,$ which specify initial data (5) for these solutions, enter the first polynomial
$S_1$. Hence,
\begin{equation}
S_0 = S_1(0) = \frac{1}{2}(1-\beta_1), \;\;\; \dot{S}_0 = S_1'(0) = \frac{\alpha_1}{2},
\end{equation}
where $\alpha_1,\; \beta_1$ have the form (20). Because $a_0, a_1$ are zeros of this polynomial, they are
related to the initial data by the following formulas
\[ a_0 =
 \frac{\dot{S}_0 - \sqrt{\Delta}}{1 - 2 S_0}, \;\; a_1 =  \frac{ \dot{S}_0 + \sqrt{\Delta}}{1 - 2 S_0},
\]
where $ \Delta = \dot{S}_0^2 - 4 S_0^2 + 2 S_0.$ By the assumption,  for the considered solutions   $ -1 < a_0
\leq 0, \;\; 0 < a_1 <1,$ see Fig. 4. It follows that
\begin{equation}
2 S_0 - \frac{1}{2} < \dot{S}_0 < \frac{1}{2} - 2 S_0, \;\;\; 0 \leq S_0 < \frac{1}{4}.
\end{equation}
These inequalities correspond to the $ S_0 \geq 0$ part of the interior of rhomb III in Fig. 1.

\subsection{ Solutions from the bottom-right open edge of rhomb III}
In the present case we take the trivial solution in the regions $y \leq 0$ and $ y \geq 1$. At the point $y=1$
the trivial solution matches the partial solution $S_-.$ Because that point lies on the characteristic line of
the signum-Gordon equation, we demand only continuity of $S(y)$. The solution $S_-$ in turn matches  a partial
solution $S_+$ at a point $y_1$ from the interval $ 0< y_1 <1.$  Here we also demand continuity of derivatives
$dS/dy$. Of course, both $S_-(y_1)$ and $S_+(y_1)$ are equal to 0.  The calculations are straighforward. It
turns out that
\begin{equation}
v_1 = \frac{1}{2S_0} -1.
\end{equation}
This formula implies that $ v_1 >1.$ The solution $\varphi$ has the following form
\begin{equation}
\varphi(x,t) = \left\{ \begin{array}{lcl}
\varphi_0 =  0  & \;\;\; \mbox{for} \;\;\; & x \leq t, \\
\varphi_1 = \frac{1}{2 (v_1 -1)} (x -t) (x - v_1 t)&     \;\;\; \mbox{for} \;\;\;   &    t \leq x \leq v_1 t, \\
\varphi_2 = (x +t) [  S_0(x+t) - \frac{t}{2}] &    \;\;\; \mbox{for} \;\;\;   &x \geq v_1 t,
\end{array} \right.
\end{equation}
see  Fig. 6.
\begin{center}
\begin{figure}[tph!]
\hspace*{1cm}
\includegraphics[height=4.5cm, width=12cm]{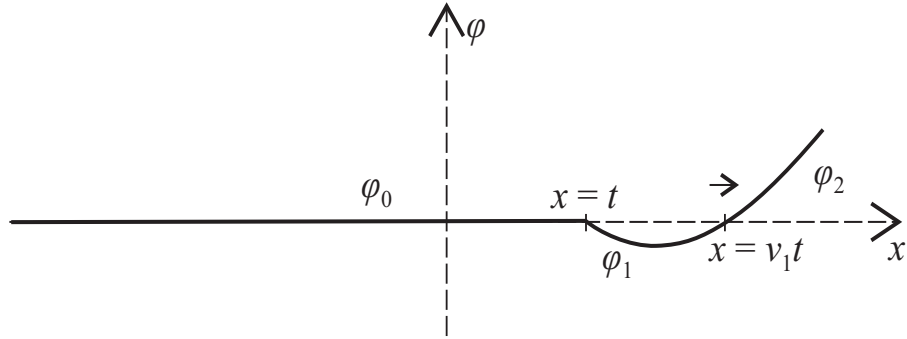}
\caption{ The self-similar solution (29).  }
\end{figure}
\end{center}

These solutions have the following initial data
\begin{equation}
\dot{S}_0 = 2 S_0 - \frac{1}{2}, \;\;\; 0 < S_0 < \frac{1}{4}.
\end{equation}
They lie on the open segment which is the interior of the  $S_0 >0, \; \dot{S}_0 <0$  edge of rhomb III in Fig.
1.

Let us note that these solutions can be regarded as a limiting case of solutions discussed in the previous
subsection obtained by putting  $a_0 =-1$ and keeping $a_1$ in the open interval $(0, 1)$.

\subsection{ Solutions with $v =1$}
 In this case we combine the trivial solution and an $S_+$ solution. They match continuously at the point $y=1$.
 \begin{equation}
\varphi(x,t) = \left\{ \begin{array}{lcl}\varphi_0 = 0 & \;\;\; \mbox{for} \;\;\; & x\leq t,      \\
 \varphi_1 =  (x-t)\:[S_0 (x-t) + t/2]  & \;\;\; \mbox{for} \;\;\; &  x > t,
\end{array}
\right.
\end{equation}
see Fig. 7.  In particular, for  $S_0 = 1/4$
we have $  \varphi(x,t) = (x^2 - t^2)/4$  for $x \geq t,$ and  $ \varphi =0$ for $x\leq t$. \\

\begin{center}
\begin{figure}[tph!]
\hspace*{1cm}
\includegraphics[height=4.5cm, width=12cm]{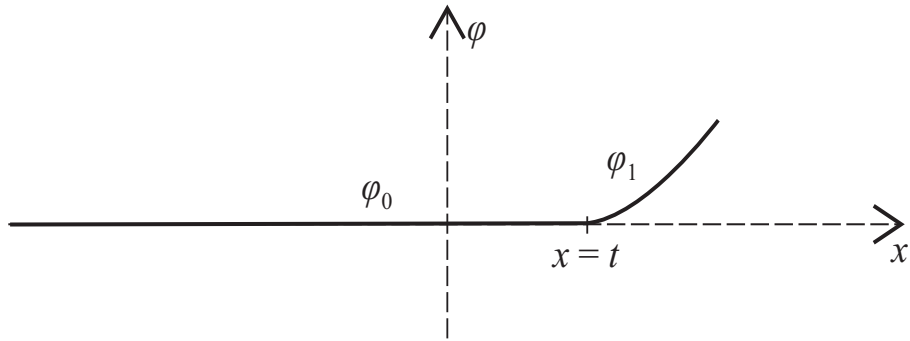}
\caption{ The self-similar solution (31). }
\end{figure}
\end{center}

The parameters $S_0, \dot{S}_0$ for these solutions obey the relation
\begin{equation}
\dot{S}_0 = - 2 S_0 + \frac{1}{2}, \end{equation} where $S_0 \geq 0.$

Some of these solutions are related to the ones discussed in subsection 3.3: if we take $a_1 = 1$ and $ -1 \leq
a_0 \leq 0, $ then only the first polynomial $S_1$ is present and it leads to the solution of the form (31).
However, the restriction $ -1 \leq a_0 \leq 0 $ implies that $ 0 \leq S_0 \leq 1/4$ because $ S_0 = - a_0/2(1-
a_0).$ This corresponds to that part of the $v=1, \; S_0 \geq 0$ half-line, see Fig. 1, which is the upper-right
edge of rhomb III.

 \section{Non self-similar solutions with a self-similar component}
In this Section we would like to present a certain interesting class of solutions of the signum-Gordon equation
which are not self-similar. The corresponding initial data for them  are different from (5): they  have the form
\begin{equation}
\varphi(x,0) = \left\{ \begin{array}{lcl} 0 & \;\;\; \mbox{for} \;\;\; & x \leq 0, \\
\alpha x  & \;\;\; \mbox{for} \;\;\; &  x \geq 0,
\end{array}
\right. \left.
\partial_t\varphi(x,t)\right|_{t=0}= 0,
\end{equation}
where $ \alpha >0$ is a free parameter.  The  solution $\varphi$ was first obtained numerically in a context
which is not relevant here. Computer simulations  showed that, rather surprisingly, certain important
quantitative characteristics of the evolution of $\varphi$ apparently did not depend on the parameter $\alpha$
at all.  Such numerical findings have motivated us to study these solutions in more detail. It has turned out
that the solutions have a self-similar component during a certain finite time interval. For this reason we would
like to present them here.

Because the signum term in Eq. (1) remains constant until $ \varphi$ becomes equal to zero, the analytic
solution of Eq. (1) with the initial data (33) can be constructed piecewise. In each interval on the $x$ axis
such that $\varphi$ has a constant sign in it, one can use the well-known formula for general solution of the
one-dimensional wave equation, suitably modified to incorporate the constant $+1$ (or $-1$) term in the
equation:
\[
\varphi(x,t) = h(x-t) + g(x+t) + c_1 t^2  - c_2 x^2,
\]
where $ 2(c_1 + c_2) = \pm 1 (= sign(\varphi)). $ The functions $ h, g$ and the constants $c_1, c_2$ are
determined from the initial conditions and from the matching conditions.   The resulting solution has a
relatively simple form when the time $t$ is not too large. The evolution of $\varphi$ can be divided into
distinct stages. In the first one time $t$ changes from 0 to $\alpha$.  Then
 \begin{equation}
\varphi_a(x,t) = \left\{ \begin{array}{lcl} \varphi_{0}(x,t) =0 & \;\;\; \mbox{for} \;\;\; & x \leq -t, \\
\varphi_{a1}(x,t)= \frac{1}{8} (x+t) (4 \alpha + x  - 3 t) & \;\;\; \mbox{for} \;\;\; & -t \leq x \leq t,
\\ \varphi_{a2}(x,t) = \alpha x - \frac{t^2}{2} & \;\;\; \mbox{for} \;\;\; & x\geq t.
\end{array}
\right.
\end{equation}
The function $\varphi_{a1}$ remains positive in the interval $x \in(-t,t]$ until $t = \alpha.$ Notice that the
border point betwen the supports of  functions $\varphi_0,\: \varphi_{a1}$ moves with the velocity $-1$ which
does not depend on the slope $\alpha$ of the initial shape (33) of $\varphi$. This fact is related to the jump
of the value of $\partial_x\varphi_a$ at the point $x = -t$ - such discontinuity is allowed for by the wave
equation only on characteristics. Hence the border point between the supports of the functions $\varphi_0,
\varphi_{a1}$ has to move with velocity -1 until $\left. \partial_x\varphi_{a1}\right|_{x=-t}$ vanishes. This
happens at the moment $t = \alpha$.

At that moment
 \begin{equation}
\varphi_a(x,\alpha) = \left\{ \begin{array}{lcl} 0 & \;\;\; \mbox{for} \;\;\; & x \leq - \alpha, \\
 \frac{1}{8} (x + \alpha)^2 & \;\;\; \mbox{for} \;\;\; & -\alpha \leq x \leq \alpha,
\\ \alpha (x - \frac{\alpha}{2}) & \;\;\; \mbox{for} \;\;\; & x\geq \alpha,
\end{array}
\right.
\end{equation}
and
 \begin{equation}
(\partial_t\varphi_a)(x,\alpha) = \left\{ \begin{array}{lcl} 0 & \;\;\; \mbox{for} \;\;\; & x \leq -\alpha, \\
- \frac{1}{4} (x + \alpha) & \;\;\; \mbox{for} \;\;\; & -\alpha \leq x \leq \alpha,
\\ - \alpha  & \;\;\; \mbox{for} \;\;\; & x\geq \alpha.
\end{array}
\right.
\end{equation}
These values of $\varphi, \: \partial_t \varphi$ constitute the initial data for the second stage of evolution
of $\varphi$ which lasts until $t = 2 \alpha$. Here the crucial observation is that $\varphi_a, \:
\partial_t \varphi_a$ at the time $\alpha$ have the shape which coincides with the self-similar initial data (5) in
which $x$ is replaced by $ \xi = x + \alpha.$ The corresponding solution belongs to the class discussed in
subsection 3.4. It has $S_0 = 1/8, \: \dot{S}_0 = - 1/4, $ and $v_1 =3$. It is interesting that the velocity
$v_1$ does not depend on the parameter $\alpha$. The $\varphi_1$ part, see Fig. 6, begins to appear at the point
$x = -\alpha$ at the time $t=\alpha$. It expands to the right with velocity $+3$, while the left end of its
support moves slower, with  velocity $+1.$  Thus, we expect that in the interval $ x \in (- \infty, t]$ the
solution has the form of time and space translated self-similar solution (29) ($t \rightarrow \tau = t - \alpha,
\; x \rightarrow \xi = x + \alpha$). At the point $x =t$ it continuously matches the function $\varphi_{a2}$
from formula (34). With these hindsights one can easily construct the analytic solution. In the time interval $
t \in [\alpha, 2 \alpha]$ we have $ \varphi(x,t) = \varphi_b(x,t),$ where \\
\begin{equation}
\varphi_b(x, t) = \left\{ \begin{array}{lcl} 0 & \;\;\; \mbox{for} \;\;\; & x \leq t - 2\alpha, \\
\varphi_{b1} = \frac{1}{4}  (x - x_0(t)) ( x - 3t + 4 \alpha) & \;\;\; \mbox{for} \;\;\; & t - 2 \alpha \leq x
\leq 3 t - 4 \alpha,
\\
\varphi_{b2} =  \frac{1}{8}  (x + t) ( x - 3t + 4 \alpha)  & \;\;\; \mbox{for} \;\;\; & 3 t - 4 \alpha \leq x
\leq  t,
\\
\varphi_{b3} = \alpha x - \frac{1}{2} t^2 & \;\;\; \mbox{for} \;\;\; & x\geq \ t.
\end{array}
\right.
\end{equation}
Here \\

\[ x_0(t) = t - 2 \alpha. \]

 Notice that $\varphi_{b3}$ describes the ``freely falling'' halfline $\varphi
= \alpha x$, which is the remnant of the initial data (33).

At the time $t = 2 \alpha$ another interesting thing happens. At this moment the support of $\varphi_{b2}$
becomes reduced to the point $ x = 2 \alpha$, and simultaneously $ \varphi_{b3}(2\alpha, 2\alpha) =0. $ The
right end of the support of $\varphi_{b1}$, which moves with the velocity $+3$, hits the support of
$\varphi_{b3}$. At this time the length of the support of $\varphi_{b1}$ is equal to $2 \alpha$.

In the next time interval, $t \in[2\alpha, 3 \alpha]$, the solution $\varphi = \varphi_c$ has the following
form, see also Fig. 8,\\

\begin{equation}
\varphi_c(x, t) = \left\{ \begin{array}{lcl} \varphi_0 = 0 & \;\;\; \mbox{for} \;\;\; & x \leq t - 2\alpha, \\
\varphi_{c1} = - \frac{1}{4}  (x - x_0(t)) ( 3t - x  - 4 \alpha) & \;\;\; \mbox{for} \;\;\; & t - 2 \alpha \leq
x \leq  4 \alpha - t,
\\
\varphi_{c2} & \;\;\; \mbox{for} \;\;\; & 4 \alpha -t \leq x \leq t,
\\
\varphi_{c3} & \;\;\; \mbox{for} \;\;\; & t \leq x \leq \frac{t^2}{2\alpha},
\\
\varphi_{c4} = \alpha x - \frac{1}{2} t^2 & \;\;\; \mbox{for} \;\;\; & x\geq \ \frac{t^2}{2 \alpha},
\end{array}
\right.
\end{equation}
where \\

\[
\varphi_{c2} = \frac{3}{8} x^2 - \frac{3}{4} x t + \frac{7}{8} t^2 + \alpha^2 - \frac{1}{2} \alpha t +
\frac{7}{2} \alpha x - \frac{\sqrt{\alpha}}{3} \left[ 2(x + t) + \alpha\right]^{3/2},
\]

\[
\varphi_{c3} = \frac{t^2}{2} + 3 \alpha x  - \frac{\sqrt{\alpha}}{3} \left[ 2(x + t) + \alpha\right]^{3/2} +
\frac{\sqrt{\alpha}}{3} \left[ 2(x - t) + \alpha\right]^{3/2} + \frac{2 \alpha^2}{3}.
\]
Notice that $\varphi_{c1}$ coincides with $\varphi_{b1}$ from formula (37). \\

\begin{center}
\begin{figure}[tph!]
 \hspace*{1cm}
\includegraphics[height=4.5cm, width=12cm]{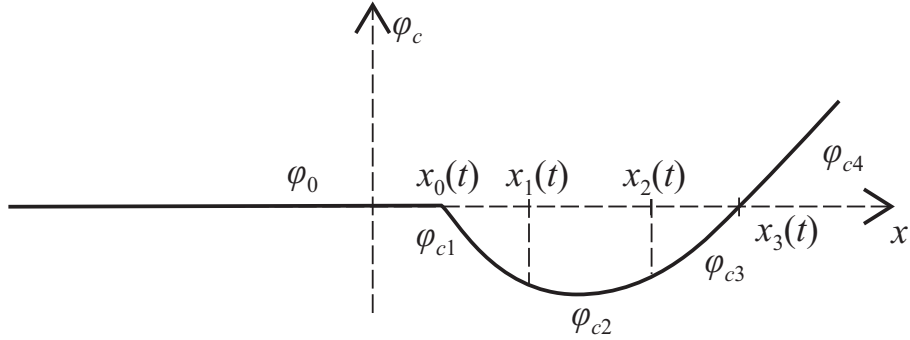}
\caption{ The solution $\varphi_c$.  Here $x_0(t) = t - 2 \alpha, \; x_1(t) = 4 \alpha -t, \; x_2(t) = t, \;
x_3(t) = \frac{t^2}{2 \alpha},$ and $t \in [2 \alpha, \: 3 \alpha].$}
\end{figure}
\end{center}
Thus, the point $x_3(t)$ (see Fig. 8) accelerates from the initial velocity $+ 2 $ to $ \dot{x}_3(3\alpha) =
9/2.$ At the time $t = 3 \alpha$ the points $ x_0, x_1$ meet each other, and the support of $\varphi_{c1}$
vanishes. It is the beginning of the fourth stage of evolution of $\varphi$.

\begin{center}
\begin{figure}
\includegraphics[height=5cm, width=13cm]{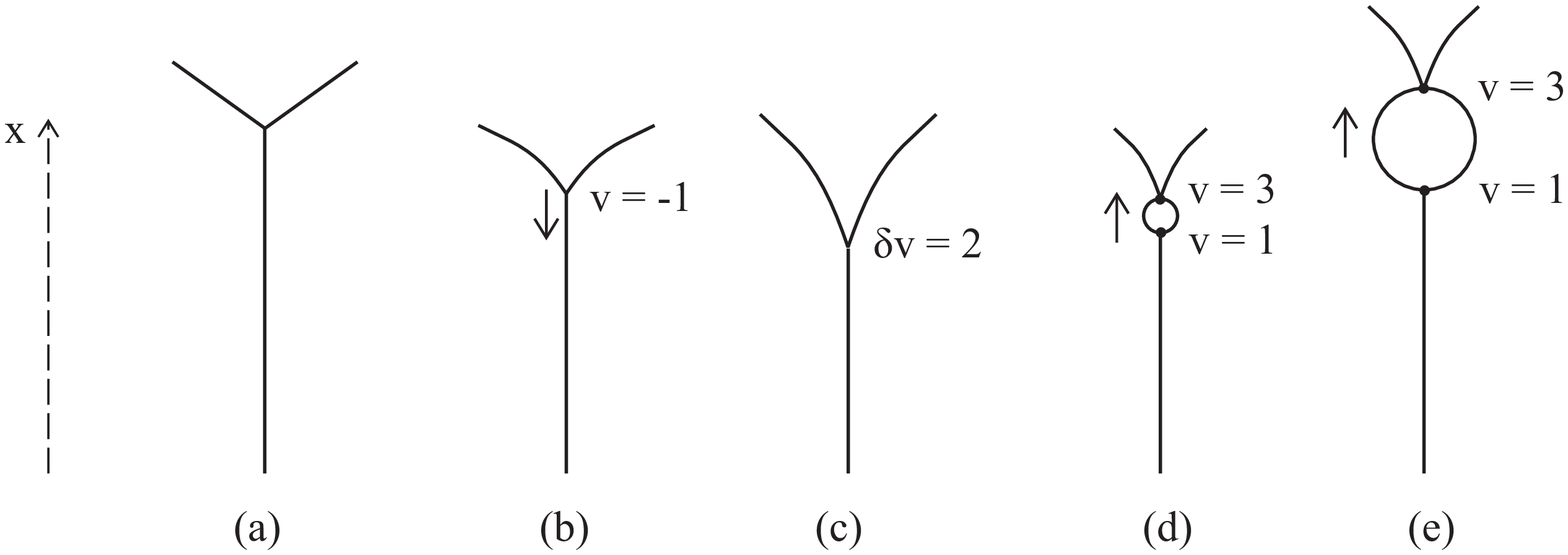}
\caption{The process of merging of two strings as described by solutions (34), (37). The 'bubble' visible in
frames (d), (e) is a part of the self-similar solution. Frame (c) shows the cusp formed at the time $T_1 =
\alpha.$  Frames (d), (e) show snapshots of the two strings at the times $t_1, t_2$ such that $\alpha < t_1 <
t_2 < 2 \alpha.$}
\end{figure}
\end{center}

When considered in the context of the interacting strings mentioned in the Introduction, the solution discussed
above provides an interesting picture of first stages of merging of the strings. In the initial state at $t =0$,
frame (a) in Fig. 9, the strings form a `Y' shape with the vertex at the point $x=0$, and the angle $\alpha$
between the two rectilinear pieces of strings in the region $x >0.$ In the region $x <0$ the strings have
already merged.  In the time interval $[0, \alpha)$ the vertex moves with the velocity -1 until the cusp is
formed, see frame (c) in Fig. 9 - this happens at the time $t=\alpha$. At this moment the bubble appears, which
grows and moves  along the strings in accordance with the self-similar solution of the signum-Gordon equation.
Such smooth evolution goes on until $t = 2 \alpha.$ In the third time interval, i.e. when $t\in [2 \alpha, 3
\alpha), $ the bubble growes and still moves along the strings, but now it is distorted by two light fronts
$x_1(t), x_2(t)$ travelling along it, as follows from the solution $\varphi_c$ given by formula (38). The
evolution can be calculated also for times $t > 3 \alpha,$ but we shall not dwell on it.

\section{ Summary and remarks }

1. The signum-Gordon equation appears in several problems of classical physics. It has the peculiar
discontinuous nonlinearity, but this does not mean that the equation is intractable. We have constructed the
full set of its self-similar solutions with initial data (5). The manifold of these solutions, presented as the
map in  Fig. 1, is amazingly rich. Equally astonishing is the fact that it has been possible to find exact
analytic forms of all solutions.

The self-similar solutions may also appear as a component of generally non self-similar solution. This
possibility is illustrated by the example presented in Section 4. It is interesting that the pattern of
evolution of $\varphi$ does not depend on the slope $\alpha$ in the initial data (33) - probably this property
reflects the scaling symmetry of the signum-Gordon equation. \\

\noindent 2. There are several obvious directions for extending our results. First, one may consider the
four-parameter set of self-similar initial data (4) instead of the two-parameter set (5). It seems that not
always the corresponding solutions can be obtained by straightforward combining solutions we have given above.
Therefore, the map of such solutions will be truly four-dimensional.

One may also be interested in stability of our solutions. For that matter, we have not seen any instability in
the numerical investigations of the self-similar solutions. This seems to indicate that unstable modes, if
present at all, grow rather slowly.

Probably the most interesting topic is related to dynamics of topological compact solitons (compactons) found in
\cite{7}. These solitons have a piece-wise parabolic shape, and the pertinent field-theoretic model contains
sectors which are described by the signum-Gordon equation. Therefore, one may expect that certain self-similar
waves will appear in such processes as scattering of the solitons, or relaxation of a single excited soliton.
Perhaps one can even provide the exact analytic description of certain stages of such processes. This would give
the much desired counterbalance to purely numerical investigations which are  dominant in the literature so far.

\section{Acknowledgements}
This work is supported in part by the Programme "COSLAB" of the European Science Foundation. H. A. gratefully
acknowledges hospitality and support during his stays at Niels Bohr Institute and at Yukawa Institute of
Theoretical Physics, where parts of this work were done.


\begin{thebibliography}{99}
\bibitem{1} H. Arod\'z, P. Klimas and T. Tyranowski, Acta Phys. Pol. \textbf{B36}, 3861 (2005).
\bibitem{2} H. Arod\'z, P. Klimas and T. Tyranowski, Phys. Rev. \textbf{E 73}, 046609 (2006).
\bibitem{3} L. Perivolaropoulos, Nucl. Phys. \textbf{B375}, 665 (1992).
\bibitem{4} G. I. Barenblatt, \emph{Scaling, selfsimilarity, and intermediate asymptotics}. Cambridge University
Press, 1996.
\bibitem{5}  L. Debnath,\emph{ Nonlinear Partial Differential Equations for Scientists and Engineers}. Birkhäuser,
Boston-Basel-Berlin, 2005. Chapter 8.
\bibitem{6} R. D. Richtmyer, \emph{Principles of Advanced Mathematical Physics}. Springer-Verlag,
New York-Heidelberg-Berlin, 1978. Section 17.3.
\bibitem{7} H. Arod\'z, Acta Phys. Pol.\textbf{ B33}, 1241 (2002).

\end{thebibliography}
\end{document}